# Teacher training in the age of AI: Impact on AI Literacy and Teachers' Attitudes

Julia Lademann [a,*], Jannik Henze [b], Nadine Honke [a], Caroline Wollny [a], and Sebastian Becker-Genschow [a]

[a]*University of Cologne, Faculty of Mathematics and Natural Sciences, Digital Education Research, 50931 Cologne, Germany*
[b]*University of Cologne, Faculty of Mathematics and Natural Sciences, Institute for Physics Education, 50931 Cologne, Germany*
[*]*Corresponding author: julia.lademann@uni-koeln.de, Heinrich-Hoerle-Straße 2, 50354 Hürth, Germany*

**Abstract.** The rapid integration of artificial intelligence (AI) in education requires teachers to develop AI competencies while preparing students for a society influenced by AI. This study evaluates the impact of an online teacher training program on German in-service teachers' AI literacy, usage behaviors, and attitudes toward AI. A pre-post design study was conducted with teachers ($N_1$ = 291 for AI literacy, $N_2$ = 436 for attitude assessment) participating in the course. The program combined synchronous and asynchronous learning formats, including webinars, self-paced modules, and practical projects. The participants exhibited notable improvements across all domains: AI literacy scores increased significantly, and all attitude items regarding AI usage and integration demonstrated significant positive changes. Teachers reported increased confidence in AI integration. Structured teacher training programs effectively enhance AI literacy and foster positive attitudes toward AI in education.

## I. INTRODUCTION

Since ChatGPT was released in 2022, there has been increasing interest in Artificial Intelligence (AI), particularly in the domain of generative AI. AI has become highly prominent in educational contexts (Fissore et al., 2024; Küchemann, Steinert, et al., 2024a; Wollny et al., 2021). The rapid pace at which AI has entered educational settings presents school systems with unprecedented challenges that differ fundamentally from those posed by previous digital tools, necessitating a swift response from the research community (Dunnigan et al., 2023).

Research has identified numerous advantages of AI use in education (Adiguzel et al., 2023; Küchemann, Avila, et al., 2024; Neumann et al., 2024). For students, AI enhances personalized learning experiences, improves accessibility for diverse learners, and provides digital learning assistance for research and writing tasks (Kasneci et al., 2023). Teachers benefit from AI support in lesson planning, generation of differentiated learning materials, writing personalized feedback, and increased efficiency in administrative tasks, allowing them to focus more on pedagogical interactions with students (Neumann et al., 2024; Seßler et al., 2023; Zhai, 2023b; Zheng et al., 2023).

To leverage these advantages effectively, teachers play a pivotal role. They are responsible not only for incorporating AI into their own teaching practice, but also for preparing students for a society increasingly influenced by AI technologies (Dolezal et al., 2025). This requires educators to develop comprehensive AI competencies – not only the capacity to use AI proficiently themselves but also the ability to teach students about AI applications while addressing both its potential benefits and risks (Fissore et al., 2024; Moura & Carvalho, 2024). However, the complexity of AI as a broader concept – resembling the internet rather than a single tool with a user manual – complicates teacher education in this domain, in contrast to the introduction of other digital tools such as calculators or specific learning platforms (Dunnigan et al., 2023).

The successful integration of AI in education depends heavily on teachers' initial perspectives and attitudes toward these technologies (Kaplan-Rakowski et al., 2023; Sugar et al., 2004). Therefore, educators must develop not only AI literacy but also positive attitudes toward AI and its educational applications. Understanding fundamental principles of AI becomes crucial for identifying potential risks and leveraging its benefits effectively (Moura & Carvalho, 2024).

Despite the demonstrated efficacy of AI tools in supporting teachers' professional practice (Neumann et al., 2024; Zhai, 2023a), research addressing teachers' attitudes toward AI and its potential for teaching practices remains limited (Nazaretsky et al., 2022). This gap highlights the need for targeted professional development programs and systematic evaluation of their effectiveness in enhancing both teachers' AI literacy and their willingness to incorporate AI tools into their teaching practice.

The present study addresses this research gap by evaluating the impact of an online training program on in-service teachers' AI literacy and their perspectives on various AI-related topics. Specifically, in addition to measuring the effect on the teachers' AI literacy, the study examines their usage behaviors (present and future), potential reductions in fears related to educational AI use, and the teachers' preparedness to incorporate AI into their lessons and lesson planning.

## II. STATE OF RESEARCH

### A. AI literacy

The term "AI literacy" is widely used and there are different approaches to conceptualizing it (Laupichler et al., 2022). The present study aligns with the established framework proposed by Long and Magerko which characterizes AI literacy not solely as the possession of fundamental knowledge about AI but rather as "a set of competencies that enables individuals to critically evaluate AI technologies; communicate and collaborate effectively with AI; and use AI as a tool online, at home, and in the workplace" (Long & Magerko, 2020). This definition reflects a shift from passive knowledge acquisition to active competency development, acknowledging that effective AI literacy requires both understanding and practical application skills. The framework encompasses 17 competencies, organized around five fundamental questions (What is AI? What can AI do? How does AI work? How should AI be used? And How do people perceive AI?).

The AI literacy test employed in this study was developed based on the aforementioned conceptualization of AI literacy, ensuring its applicability across diverse educational context and participants backgrounds (Hornberger et al., 2023).

It is important to note that AI literacy exists within a broader ecosystem of digital competencies. While closely related to concepts such as digital literacy and computational thinking, AI literacy specifically addresses the unique challenges and opportunities presented by intelligent systems that can adapt, learn, and make autonomous decisions.

Research suggests that educators play an important role in fostering AI literacy among learners (Dunnigan et al., 2023). This responsibility necessitates that educators themselves possess a certain degree of AI literacy to effectively integrate these technologies into their pedagogical practice and their curriculum while maintaining critical perspectives on their use.

### B. AI in Education

**a. Application, Potentials, and Risks**

AI has the potential to transform the traditional way of teaching and learning, leading to a growing interest in research concerning the domain of AI in education (Celik et al., 2022; Popenici & Kerr, 2017). Studies demonstrate both acceptance and utilization of AI in educational settings (Chen et al., 2020; Fissore et al., 2024), with school leaders recognizing generative AI's potential for their institutions (Dunnigan et al., 2023). Educators acknowledge that generative AI could enhance their professional development and reduce their workload while serving as a valuable tool for students (Cojean et al., 2023; Kaplan-Rakowski et al., 2023; Ng et al., 2025).

Despite this promising outlook, the implementation of these practices remains limited. A 2025 study revealed that less than half of the surveyed teachers currently employ AI tools in their daily practice (Cheah et al., 2025). The findings indicate that teachers primarily employ AI applications for lesson planning, creating exercises, and adapting content for diverse student groups, followed by its usage for grading and assessment.

In contrast, student adoption exceeds teacher usage. While teachers have not yet actively promoted student engagement with AI tools (Galindo-Domínguez, Delgado, Losada, et al., 2024), students themselves are already using these technologies frequently (Higgs & Stornaiuolo, 2024; Kotchetkov & Trevor, 2025). A study conducted at European schools shows 74% of students believe AI will play an important role in their future careers, and 66% believe that access to AI is significant for their success in school. Among AI tools, 48% of European students identify ChatGPT as their primary choice (Vodafone Stiftung, 2025).

AI chatbots as personal assistant provide learners with individualized feedback and adapt to specific learning difficulties by addressing student questions (Adiguzel et al., 2023; Kasneci et al., 2023; Neumann et al., 2024). It has been demonstrated that these tools enhance student performance (Wu & Yu, 2024) and promote productivity and motivation (Fauzi et al., 2023; Küchemann, Avila, et al., 2024). Notably, learning with explanations generated by an AI chatbot fosters situational interest, positive emotions, and self-efficacy expectations while reducing cognitive load (Lademann et al., 2025).

The high number of students already using AI suggests that they are not averse to incorporating it into academic and daily life. However, there is a possibility that they may possess a deficiency in knowledge regarding ethical considerations and safe utilization of AI tools. Survey results show only 30% of students report being "somewhat familiar" with AI, while merely 11% indicated being "very familiar" (Vodafone Stiftung, 2025). This highlights the necessity of educating teachers about responsible AI utilization and associated risks, including data protection concerns, AI biases, deepfakes, and AI hallucination (Akgun & Greenhow, 2022; Karan & Angadi, 2023; Krupp et al., 2024; Küchemann, Steinert, et al., 2024b).

**b. The Role of Educators with regard to AI**

Educators face a dual responsibility: developing their own AI competency for effective interaction with these tools while empowering students to engage safely and responsibly with AI technologies. This is amplified by the observation that technology integration methods matter more than technology itself (Cheah et al., 2025; Crawford et al., 2023; OECD, 2015), underlining educator's crucial role in incorporating novel digital tools into the classroom environment and guiding students for future digital landscapes.

Teachers must develop familiarity with fundamental concepts related to AI, computational thinking, and machine learning principles (Asunda et al., 2023). In addition, teachers must possess the ability to comprehend various AI applications and identify appropriate tools for specific contexts across diverse grade levels, curricula, and subject areas (Cheah et al., 2025).

However, research indicates that teachers are not adequately prepared to integrate generative AI into daily practice (Cheah et al., 2025). Notably, 68 % of students believe teacher AI competency depends on chance rather than systematic preparation (Vodafone Stiftung, 2025), underscoring the urgent need for increased training opportunities in the field of AI literacy.

Understanding how to address these preparation gaps requires examining the relationship between knowledge and attitudes. Research reveals a positive correlation between teachers' AI literacy and their attitudes toward AI technologies (Galindo-Domínguez, Delgado, Campo, et al., 2024; Hornberger et al., 2023). Teachers with greater AI knowledge and readiness tend to view these technologies more favorably, suggesting that familiarity reduces apprehension (X. Wang et al., 2023). This finding implies that professional development programs should focus on building both technical competencies and confidence through hands-on experience with AI tools.

## C. Teacher Training Programs

The integration of generative AI into educational practice is an impending change requiring collective effort to navigate. Unlike established technologies, AI does not possess a set of established guidelines or a comprehensive manual. Moreover, educators are already confronted with an extensive workload. New digital initiatives in educational settings, particularly in schools, encounter resistance to change, hindering implementation (Dunnigan et al., 2023; Stacey et al., 2023).

Given the insights outlined in **II.B.b**, targeted support for educators for AI integration into their daily pedagogical practices becomes imperative (Bergdahl & Sjöberg, 2025). Teacher training programs represent a valuable opportunity for in-service teacher education about new technologies and addressing academic resistance to the adoption of new technologies (Nazaretsky et al., 2022; Rienties, 2014). Historical patterns show that inadequate teacher training and preparation engenders apprehension and anxiety regarding the integration of technologies in practical applications (Ally, 2019; Kaplan-Rakowski et al., 2023; T. Wang & Cheng, 2021; Yang & Chen, 2023).

As educators become increasingly exposed to AI, they may develop more confidence and motivation to incorporate AI into teaching practices. Teacher attitudes toward artificial intelligence positively correlate with their frequency of utilization of generative AI. (Kaplan-Rakowski et al., 2023; Kuleto et al., 2022). Teacher training programs, such as the one evaluated in this study, facilitate controlled integration of new technologies within secure environments and under expert guidance, encouraging the utilization of AI tools while providing feedback and opportunities for self-reflection.

Educators demonstrate profound interest in acquiring AI training within educational contexts (Fissore et al., 2024; Ng et al., 2025). However, this interest is not merely superficial; rather, it signifies a critical need to enhance understanding of AI technologies and cultivate practical implementation capacity (Fissore et al., 2024; Lo, 2023). Effective AI training programs should integrate AI literacy with comprehensive understanding of existing AI tool landscapes to prepare teachers for varied scenarios (Galindo-Domínguez, Delgado, Losada, et al., 2024).

Despite this clear need, there is a lack of research on professional teacher training (Sanusi et al., 2023). The objective of the present study is to evaluate whether teacher training programs can enhance AI literacy and whether participants subsequently feel prepared to integrate AI technologies into their teaching. Furthermore, participation's influence on the utilization of AI by the participants and their assessment of future opportunities and risks is examined. This may indicate whether in-service teachers develop confidence in the utilization of generative AI, thereby reducing fears while preparing them to incorporate AI into daily work and student instruction.

## III. RESEARCH QUESTIONS

The current state of research implies that further studies in this still new and under-researched area are both useful and imperative. Teacher training programs present an opportunity to educate in-service teachers about new technologies while addressing existing academic resistance and reducing anxiety regarding the integration of these technologies in educational contexts. However, further inquiry is necessary to gain a more comprehensive understanding of the impact of specific teacher training programs regarding AI. As previously stated in **II.B.b**, it has been demonstrated that AI literacy and positive attitudes toward AI technologies have a significant impact on the preparedness and the willingness of teachers to incorporate AI into their daily teaching practice. Nonetheless, at this juncture, the question remains as to whether training programs on AI in general and the examined training program in particular can positively influence the aforementioned aspects. In this study, an evaluation of a teacher training program for in-service teachers of all subjects and school forms in Germany was conducted (The structure of the training program is delineated in **IV.A.c.**). The objective of the present study is to address the following key research questions:

How does participation in the training program affect the participants'

> RQ1: … AI literacy, both generally and within specific competence domains?
>
> RQ2: … current use and intended integration of AI both in their daily lives and in an educational context?
>
> RQ3: … way of assessing future opportunities and risks of AI in an educational context?

# IV. METHODS

## A. Study Design

### a. Sample

A total of $N_1 = 291$ teachers (male = 60, female = 231, average age 44.08 (SD 9.92)) completed the questionnaire on AI literacy and $N_2 = 436$ (male = 86, female = 348, divers = 2, average age 45.07 (SD 9.68)) teachers completed the questionnaire on attitudes toward AI. The teachers who completed the questionnaire on AI literacy were also included among those who completed the one on attitudes. The teachers were employed in a variety of types of schools throughout Germany. In accordance with standard ethical practice, they were asked to give their informed consent before completing the questionnaires, which were provided via a link. Ethical approval was not required for the study involving humans because research was conducted in accordance with local law. No personal data was collected.

### b. Procedure

The present study was conducted in a pre-post design and its objective was to evaluate a training course for teachers in the field of artificial intelligence. The course was repeated on three occasions, with a different set of participants at each iteration. The 2024 dates for the event were in January/February and May/June, and the 2025 date was February/March. In each cycle, the participating teachers completed a questionnaire once before and once after the training. In 2024, the initial two iterations of the questionnaires captured both AI literacy and respondents' attitudes regarding the subject of AI. During the last assessment in 2025, only attitudes were measured. This decision was made in light of the substantial and sufficient amount of data on AI literacy that had already been collected from the preliminary two rounds.

### c. Structure of the Training Program

In order to address the growing demand for foundational AI competencies in the education sector, the digital professional development platform fobizz[1] offers a four-week certificate course entitled "Artificial Intelligence in Everyday School and Classroom Practice." While the program was principally designed for in-service teachers in primary and secondary education, it also attracted participants from adult education, higher education, and school administration. The objective of the course was to provide foundational knowledge for the pedagogically efficacious and reflective use of artificial intelligence in school and classroom contexts. The course structure combined synchronous and asynchronous learning formats. The program's mandatory components included a kickoff webinar and the completion of a core self-paced module, titled "ChatGPT & fobizz AI in Your Classroom". During the asynchronous phase, participants had the option to engage in supplementary training units, which addressed subjects such as machine learning, AI ethics, inclusive education with AI, and data protection. Complementary live webinars and workshops were offered to provide more in-depth insights into specific educational applications of AI. During the project phase, participants developed an AI-supported lesson plan (duration: 45–90 minutes) using tools such as the fobizz AI Assistant. The final lesson plans were distributed via a digital platform designed to facilitate peer exchange and collaborative utilization of teaching resources. Upon successful completion of the program, which included participation in the kickoff session, completion of the core module, and submission of the final project, participants received a certification of 20 hours of professional development credit. Supplementary Figure 1 provides further details on the course structure.

### d. Data Collection

*AI Literacy.* A validated questionnaire was utilized to document the current level of knowledge of the participating teachers and to evaluate the impact of the training on this knowledge (Hornberger et al., 2023). The validated questionnaire consists of 29 multiple-choice items[2] and one sorting item. These items are designed to assess 14 distinct competencies of AI literacy (Long & Magerko, 2020). Each multiple-choice item has four potential responses, of which 1-4 are considered correct. The participants completed the questionnaire and assessed their level of knowledge on the topic of AI once before and once after the teacher training.

---

[1] https://fobizz.com/de/
[2] Item number 7 was eliminated during the validation process by Hornberger et al..

*Attitudes to the subject of AI.* The teachers attitudes were also measured once before and once after the participation in the training program. For this purpose, a questionnaire was developed. The questionnaire contains eight independent items and utilizes a 4-point Likert scale to assess teachers' agreement. The items were grouped into two categories to address the research questions RQ2 and RQ3 posed in this study (**Table 1**).

| No. | Item | Assigned to |
|---|---|---|
| I1 | *I regularly use AI in my daily life.* | RQ2 |
| I2 | *I use AI to help create my lessons and teaching materials.* | RQ2 |
| I3 | *I use AI to support assessments or grading.* | RQ2 |
| I4 | *AI in schools is more of a threat than an opportunity.* | RQ3 |
| I5 | *I can imagine integrating AI into my daily teaching practice.* | RQ2 |
| I6 | *I believe that AI offers an opportunity to support students.* | RQ3 |
| I7 | *AI will replace teachers in the future.* | RQ3 |
| I8 | *I feel well-prepared to integrate AI technologies into my teaching.* | RQ2 |

*Table 1: Items and assignment to research question 2 and 3.*

**B. Data Analysis**

The subsequent section outlines the analytical procedures used to address the research questions. All analyses were conducted using R version 4.3.3 (R Core Team, 2024).

Prior to analysis, the dataset was examined for completeness. Participants who did not complete both pre- and post-tests were excluded to ensure the reliability of paired comparisons. AI literacy competency scores were calculated by summing individual item scores within each competency area, while attitude and belief items were analyzed at the individual item level. The overall AI literacy score was computed as the sum of all competency scores.

Data distribution normality was assessed using the Shapiro-Wilk test with a significance level of $\alpha = 0.05$ (R package "stats", function "shapiro.test") (R Core Team, 2024). The overall AI literacy score, individual competency scores and attitude items violated normality assumptions, necessitating the implementation of non-parametric methods for all variables (see Supplementary Figure 2 and 3).

To examine significant differences between pre- and post-test measurements, Wilcoxon signed-rank tests were utilized (R package "stats", function "wilcox.test") (R Core Team, 2024), as this test is standardized for paired observations with non-normal distributions. All tests were conducted with a significance level of $\alpha = 0.05$.

Effect sizes were estimated using Cohen's d for all comparisons (R package "psych", function "cohen.d") (Revelle, 2025). Cohen's d values were interpreted according to conventional benchmarks: small ($d < 0.5$), medium ($d < 0.8$), and large ($d \geq 0.8$) effects.

## V. RESULTS

This section presents the findings from the pre- and post-test assessments, organized according to the research questions.[3]

### A. AI Literacy

The first research question (**RQ1**) examined changes in participants' overall AI literacy and across multiple competency domains.

Participants demonstrated a significant improvement in overall AI literacy from pre- to post-test. The mean score increased from 13.15 (SD 5.46) at pre-test to 15.13 (SD 5.14) at post-test ($p < 0.001$, $d = 0.37$, small effect size) (**Fig. 1**).

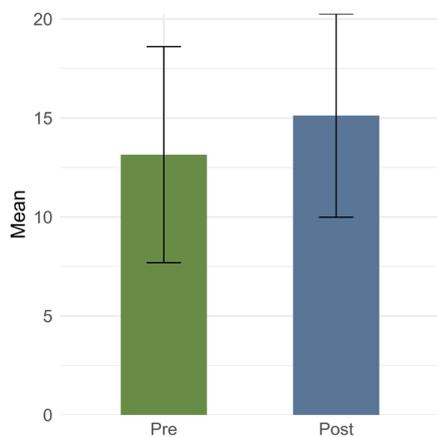

*Figure 1: Illustration of the distribution of overall AI literacy scores at pre- and post-test, showing the shift toward higher scores following the intervention.*

In order to provide a more detailed understanding of the changes in AI literacy, each competency was analyzed separately. **Table 2** presents the descriptive statistics for all measured competencies, including mean scores and standard deviations for both assessment points, along with Wilcoxon test p-values and Cohen's d effect sizes for significant results. Mean values represent the number of points achieved in each competency, with one point awarded per correctly answered item.

---

[3] For the complete results of the Wilcoxon signed-rank test see Supplementary Figure 2 and 3.

| Competency (with number of Items) | Mean (SD) Pre | Mean (SD) Post | p-value | Cohen's d |
|---|---|---|---|---|
| **C01** (2) Recognizing AI | **0.80** (0.65) | **0.91** (0.63) | 0.005** | 0.17 |
| **C02** (2) Understanding Intelligence | **1.60** (0.66) | **1.67** (0.60) | 0.093 | - |
| **C03** (2) Interdisciplinarity | **0.77** (0.76) | **0.97** (0.75) | < .001*** | 0.27 |
| **C04** (2) General vs. Narrow | **0.69** (0.74) | **0.85** (0.77) | < .001*** | 0.20 |
| **C05** (2) AI's Strengths and Weaknesses | **0.69** (0.70) | **0.86** (0.71) | < .001*** | 0.23 |
| **C07** (2) Representations | **0.46** (0.64) | **0.63** (0.68) | < .001*** | 0.26 |
| **C08** (3) Decision-Making | **0.86** (0.86) | **1.09** (0.93) | < .001*** | 0.25 |
| **C09** (3) ML Steps | **0.87** (0.89) | **1.15** (1.01) | < .001*** | 0.30 |
| **C10** (2) Human Role in AI | **0.53** (0.73) | **0.67** (0.74) | 0.007** | 0.19 |
| **C11** (1) Data Literacy | **0.47** (0.50) | **0.51** (0.50) | 0.300 | - |
| **C12** (2) Learning from Data | **1.13** (0.70) | **1.19** (0.64) | 0.240 | - |
| **C13** (1) Critically Interpreting Data | **0.80** (0.40) | **0.85** (0.36) | 0.044* | 0.13 |
| **C16** (5) Ethics | **2.67** (1.38) | **2.93** (1.26) | 0.001** | 0.20 |
| **C17** (1) Programmability | **0.80** (0.40) | **0.85** (0.36) | 0.076 | - |

*Table 2: Descriptive statistics, p-values, and effect sizes for AI literacy competencies. *p < .05, **p < .01, ***p < .001. Cohen's d effect sizes are reported only for significant results.*

The results revealed that ten of fourteen competencies showed statistically significant improvements with small effect sizes (**Fig. 2**). While competencies C02, C11, C12 and C17 did not reach statistical significance, all demonstrated positive trends with increased mean scores.

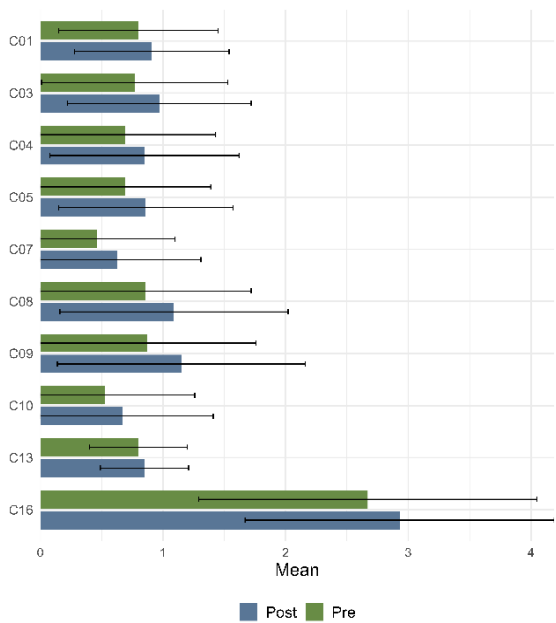

*Figure 2: Illustration of mean competency scores for AI-literacy competencies with significant changes at pre- and post-test, showing increases across targeted competencies following the intervention.*

### B. Attitudes toward AI in education

The second and third research questions investigated changes in participants' attitudes and beliefs regarding AI in educational contexts.

Participants' attitudes toward using and integrating AI in their teaching practice (**RQ2**) were assessed using five items measured on a 4-point Likert scale (0 = strongly disagree, 3 = strongly agree).

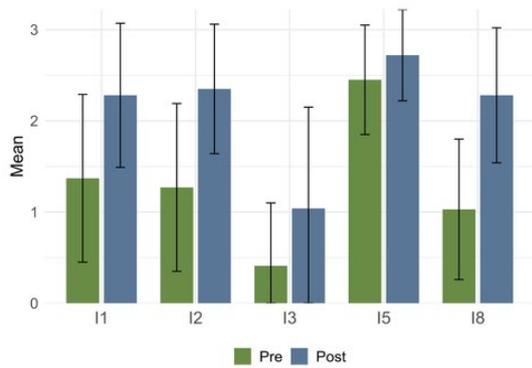

Figure 3: Presentation of the mean scores at pre- and post-tests, visualizing the magnitude of change for each item.

All five items demonstrated statistically significant improvements from pre- to post-test, with effect sizes ranging from medium (d = 0.49) to very large (d = 1.65) (**Fig. 3**). The most substantial change was observed for item **I8**, indicating that teachers feel well-prepared to integrate AI technologies into their teaching after completing the program. The descriptive statistics and inferential test results for all attitude items are displayed below (**Table 3**).

| Item | mean (SD) | | p-value | Cohen's d |
|---|---|---|---|---|
| | **Pre** | **Post** | | |
| **I1** | 1.37 (0.92) | 2.28 (0.79) | < 0.001*** | 1.07 |
| **I2** | 1.27 (0.92) | 2.35 (0.71) | < 0.001*** | 1.32 |
| **I3** | 0.41 (0.69) | 1.04 (1.11) | < 0.001*** | 0.67 |
| **I5** | 2.45 (0.6) | 2.72 (0.5) | < 0.001*** | 0.49 |
| **I8** | 1.03 (0.77) | 2.28 (0.74) | < 0.001*** | 1.65 |

Table 3: Changes in attitudes toward using and integrating AI (RQ2).

**RQ3** examined participants' beliefs about AI's role and impact in educational settings. Three items assessed these beliefs using the same 4-point scale.

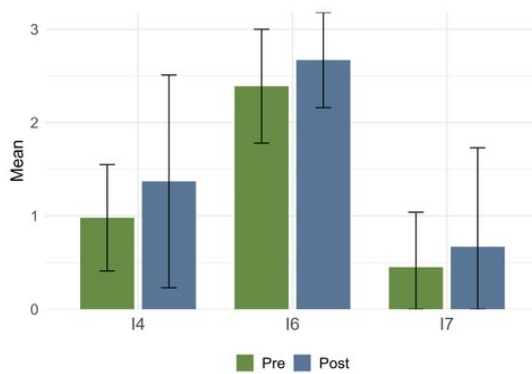

Figure 4: Display of the changes in belief scores from pre- to post-test for each item.

All three belief items showed statistically significant changes following the intervention. Effect sizes range from small (d = 0.25) to medium (d = 0.50) (**Fig. 4**), with the strongest effects observed for item **I6**, suggesting the belief that "I believe that AI offers an opportunity to support students" was most responsive to the intervention. The corresponding descriptive and inferential statistics are presented below (**Table 4**).

| Item | mean (SD) Pre | Post | p-value | Cohen's d |
|---|---|---|---|---|
| I4 | 0.98 (0.57) | 1.37 (1.14) | < 0.001*** | 0.42 |
| I6 | 2.39 (0.61) | 2.67 (0.51) | < 0.001*** | 0.5 |
| I7 | 0.45 (0.59) | 0.67 (1.06) | < 0.001*** | 0.25 |

*Table 4: Changes in beliefs about AI's role in education (RQ3).*

## VI. DISCUSSION

The present study examined the effects of a teacher training program on participants' AI literacy, utilization of AI tools, and attitudes toward AI in educational contexts. The results demonstrate significant improvements across all three investigated areas, confirming the importance of targeted professional development for successful AI integration in educational practice.

**RQ1: Effect on participants' AI literacy**

Regarding overall AI literacy, participants demonstrated a significant improvement from pre- to post-test. Prior to the implementation of the training program, teachers correctly answered approximately 13 of 30 questions on the AI literacy test. Following the training program, the number of correct answers increased to 15. Although modest effect sizes were identified, the results show a significant increase in overall AI literacy following the completion of the training program. Moreover, of the 14 evaluated competency areas, 10 demonstrated significant improvements, including domains particularly relevant to educational practice: "Recognizing AI", "AI's Strengths & Weaknesses", "Human Role in AI", and "Ethics". This suggests that the training program effectively addressed multiple dimensions of AI literacy rather than focusing narrowly on technical skills.

Nevertheless, the small effect sizes warrant careful interpretation. The validated test employed in the present study was based on a multifaceted AI literacy framework that potentially exceeded the specific educational content covered in the teacher training program. For instance, the "Ethics" category included items about „Predictive Policing", which, while important for general AI literacy, lacks a direct connection to pedagogical practice. This incongruity between the assessment tool and the training content may have served to diminish the observed effects. Notwithstanding this limitation, the significant improvements demonstrated in the majority of the competency areas underscore the program's effectiveness.

However, the fact that participants were able to solve only half of the test questions post-training reveals the existence of persistent knowledge gaps. This finding indicates the necessity for more extensive or specialized AI education programs, potentially with curricula more closely aligned to educational contexts. Subsequent iterations may benefit from the development of education-specific AI literacy assessments that more accurately reflect the unique competencies teachers require for integrating AI into the classroom.

**RQ2: Effect on current use and intended integration of AI**

The study reveals significant changes in teachers' practical utilization and intended integration of AI. The increased agreement with three key items, "I regularly use AI in my daily life.", "I use AI to help create my lessons and teaching materials." and "I use AI to support assessments or grading.", indicates immediate change in behavior.

These findings are particularly significant given the discrepancy between the recognized potential of AI (Celik et al., 2022; Kaplan-Rakowski et al., 2023) and its actual implementation in education (Cheah et al., 2025). The training program has been demonstrated to effectively address this implementation gap by offering practical, hands-on experience with AI tools in pedagogically relevant contexts, complementing theoretical knowledge.

Following the program, teachers are increasingly incorporating AI into their practices, not only in their daily work but also in the creation of lessons and learning materials, as well as in assessments and grading.

Particularly noteworthy are the significantly increased agreements to the items "I can imagine integrating AI into my daily teaching practice." and "I feel well-prepared to integrate AI technologies into my teaching.". The latter finding is crucial as it pertains both the motivational and competence dimensions of technology adoption. This dual enhancement – an increased willingness coupled with a perceived preparedness and confidence – suggests that the training program successfully addressed both psychological and practical barriers to AI adoption.

These results align with previous research demonstrating positive correlations between professional development and teacher confidence with emerging technologies (Kaplan-Rakowski et al., 2023; Kuleto et al., 2022). The program's apparent success in fostering both competence and confidence directly addresses the inadequate teacher preparation for generative AI use (Cheah et al., 2025). The underlying strategy is presumed to entail a combination of factors, namely, the acquisition of knowledge, the development of skills through practical application, and the cultivation of confidence through experimentation in a low-stakes environment.

**RQ3: Effect on assessment of future opportunities and risks of AI**

The results for RQ3 indicate a modest rise in mean values of fears and concerns regarding AI in educational systems. Despite the relatively low levels of agreement observed both before and after the program, a modest increase in agreement with the item "AI in schools is more of a threat than an opportunity." and a heightened apprehension of being replaced by AI was evident. Upon initial consideration, these results do not seem to align with previous research on teacher training programs having a positive impact in this manner (Ally, 2019; Yang & Chen, 2023). However, it is important to note that the slight increase in critical views does not inherently carry a negative connotation. The findings may also be indicative of the participants' enhanced capacity to accurately assess the risks and hazards of AI. This is an important aspect when it comes to the school context. At this juncture, conducting an extensive data analysis could be advantageous in order to examine the manner in which agreement with the two items relating to anxiety has evolved.

Nonetheless, the findings indicate a simultaneously increase in the conviction that AI represents an opportunity to support students. This suggests that the training program effectively moderated existing fears while actively fostering constructive visions for the implementation of AI in teaching. This is particularly relevant given the resistance to adopt novel digital initiatives within the educational sector (Dunnigan et al., 2023).

The findings of the present study carry significant implications for the design of future teacher professional development in AI. The efficacy of the intervention was demonstrated in addressing all three research questions, leading to notable enhancements in both AI literacy and teachers' attitudes toward AI. This underscores the necessity of systematic, comprehensive training opportunities (Fissore et al., 2024; Sanusi et al., 2023). The integration of knowledge transfer, practical application, and critical reflection proved particularly efficacious for developing both competencies and constructive attitudes.

**Limitations**

However, the study presents several limitations that constrain the generalizability and interpretation of the findings. Firstly, it remains open to question if the results of this study align with previous research on the connection between AI literacy and positive attitudes toward AI (Galindo-Domínguez, Delgado, Campo, et al., 2024; Hornberger et al., 2023). While both AI literacy and attitudes improved, the study did not examine the correlation between these changes. In addition to the enhancement of AI literacy, the positive shift in attitudes may have been influenced by access to diverse AI tools in a secure learning environment or through peer exchange.

Secondly, participant self-selection must be considered. The voluntary participation likely attracted teachers with pre-existing interest in AI and suggests motivation driven by personal interest, a higher level of openness to technological innovation, and potentially higher digital competencies. This limitation restricts the generalizability of the results, as it is possible that the participants were more receptive to the training content and more likely to implement the learned concepts. Subsequent research should investigate the impact of training programs on educators who are skeptical or resistant to AI technologies.

Thirdly, long-term data on observed effect sustainability is lacking. As data was collected immediately after the program's completion, it was not possible to ascertain whether increased AI use and positive attitudes persist over time, particularly when teachers encounter real-world implementation challenges without ongoing support. The efficacy could be further validated through longitudinal studies that would assess the translation of initial changes in practice into sustained behavioral modifications. Furthermore, a more thorough examination of the impact on fears and worries is necessary to assess and understand the changes in participant agreement with items **I4** and **I7**.

Fourthly, the study did not examine whether and how teachers transfer their acquired knowledge to students, an aspect that is particularly relevant given the high student usage of AI tools. Understanding how teachers engage in discourse regarding AI with their students and cultivate students' AI literacy represents an essential next research step.

In addition, the program's general design, which is inclusive of all subjects and school types, may have constrained its efficacy within specific teaching contexts. Future programs may benefit from more differentiated approaches tailored to specific educational contexts. The absence of oversight regarding participants' self-directed learning and elective module selections further complicates the evaluation of outcomes.

It is imperative to acknowledge that the study was conducted within a distinct cultural and educational context. Cross-cultural validation would serve to substantiate claims regarding the efficacy of training programs and identify culturally specific factors that influence the adoption of AI in educational settings.

## VII. CONCLUSION AND OUTLOOK

This study provides valuable evidence that teacher training programs effectively facilitate the integration of AI in educational settings. The significant improvements in AI literacy, practical application, and attitudes underscore that structured training programs can play a key role in preparing teachers for the digital transformation of educational practice. By addressing the current discrepancy between AI's educational potential and its practical integration in classroom settings, such programs have the potential to transform teachers from passive observers to active participants in fostering AI-enhanced learning environments.

Subsequent to the completion of the teacher training program, participants demonstrated increased AI utilization in both their daily lives and educational practice. Additionally, they reported feeling better prepared to integrate AI into their teaching, and they developed more positive attitudes regarding AI's potential benefits while maintaining awareness of associated risks. These results underscore the conclusion that teacher training programs represent a promising solution for educating teachers and preparing them for an educational system increasingly shaped by AI technologies.

This study identifies, several significant avenues for future research. First, further investigation is needed to evaluate different types of training programs and identify the most suitable approaches for various use cases and educational contexts. While the present study examined a general, cross-curricular training program, future research could explore more specialized, subject-specific training modules to better address the diverse needs of educators across different disciplines. STEM educators might benefit from AI-assisted data analysis and simulation tools while in contrast, language teachers might prioritize natural language processing applications designed to support writing and facilitate language learning. It is possible that elementary educators will place a greater emphasis on AI tools for the purposes of differentiated instruction and learning analytics than secondary teachers.

A critical priority for future research involves the development and validation of AI literacy assessment tools specifically adapted to educational contexts. Such instruments should align with established competency frameworks, for instance the UNESCO AI competency framework for teachers (UNESCO, 2024), and incorporate the competencies identified in the present study while tailoring assessment questions to educational scenarios. The field would benefit from investigating whether validated, education-specific AI literacy tests currently exist or require development.

Furthermore, research should examine the relationship between AI literacy and attitudes toward AI, particularly how teachers with initially negative attitudes respond to training interventions. A comprehensive understanding of these dynamics could inform more targeted approaches for different educator populations.

Another imperative research direction involves the assessment of the long-term stability of training effects. To ascertain the sustainability of the observed enhancements in AI literacy, usage patterns, and attitudes, longitudinal studies are necessary. These studies should examine whether improvements remain stable over time or require periodic reinforcement through continuing education initiatives.

As educational systems worldwide grapple with AI integration, the findings of this study underscore that investing in systematic teacher preparation constitutes a cornerstone of successful educational innovation, ensuring responsible and effective AI incorporation in education.

## ACKNOWLEDGEMENTS

We would like to express our sincere thanks to the fobizz team and to all participating teachers for contributing to our study.

## DISCLOSURE STATEMENT

No potential conflict of interest was reported by the authors.

This research did not receive any specific grant from funding agencies in the public, commercial, or not-for-profit sectors.

## REFERENCES

Adiguzel, T., Kaya, M. H., & Cansu, F. K. (2023). Revolutionizing education with AI: Exploring the transformative potential of ChatGPT. *Contemporary Educational Technology*, *15*(3), ep429. https://doi.org/10.30935/cedtech/13152

Akgun, S., & Greenhow, C. (2022). Artificial intelligence in education: Addressing ethical challenges in K-12 settings. *AI and Ethics*, *2*(3), 431–440. https://doi.org/10.1007/s43681-021-00096-7

Ally, M. (2019). Competency Profile of the Digital and Online Teacher in Future Education. *The International Review of Research in Open and Distributed Learning*, *20*(2). https://doi.org/10.19173/irrodl.v20i2.4206

Asunda, P., Faezipour, M., Tolemy, J., & Engel, M. (2023). Embracing Computational Thinking as an Impetus for Artificial Intelligence in Integrated STEM Disciplines through Engineering and Technology Education  *Journal of Technology Education*, *34*(2). https://doi.org/10.21061/jte.v34i2.a.3

Bergdahl, N., & Sjöberg, J. (2025). Attitudes, perceptions and AI self-efficacy in K-12 education. *Computers and Education: Artificial Intelligence*, *8*, 100358. https://doi.org/10.1016/j.caeai.2024.100358

Celik, I., Dindar, M., Muukkonen, H., & Järvelä, S. (2022). The Promises and Challenges of Artificial Intelligence for Teachers: A Systematic Review of Research. *TechTrends ; Volume 66, Issue 4, Page 616-630 ; ISSN 8756-3894 1559-7075*. https://doi.org/10.1007/s11528-022-00715-y

Cheah, Y. H., Lu, J., & Kim, J. (2025). Integrating generative artificial intelligence in K-12 education: Examining teachers' preparedness, practices, and barriers. *Computers and Education: Artificial Intelligence*, *8*, 100363. https://doi.org/10.1016/j.caeai.2025.100363

Chen, L., Chen, P., & Lin, Z. (2020). Artificial Intelligence in Education: A Review. *IEEE Access*, *8*, 75264–75278. https://doi.org/10.1109/ACCESS.2020.2988510

Cojean, S., Brun, L., Amadieu, F., & Dessus, P. (2023). Teachers' attitudes towards AI: what is the difference with non-AI technologies? *Proceedings of the Annual Meeting of the Cognitive Science Society*, *45*.

Crawford, J., Vallis, C., Yang, J., Fitzgerald, R., O'Dea, C., & Cowling, M. (2023). Editorial: Artificial Intelligence is Awesome, but Good Teaching Should Always Come First. *Journal of University Teaching and Learning Practice*, *20*(7), Article 7. https://doi.org/10.53761/1.20.7.01

Dolezal, D., Motschnig, R., & Ambros, R. (2025). Pre-Service Teachers' Digital Competence: A Call for Action. *Education Sciences*, *15*(2), 160. https://doi.org/10.3390/educsci15020160

Dunnigan, J., Henriksen, D., Mishra, P., & Lake, R. (2023). "Can we just Please slow it all Down?" School Leaders Take on ChatGPT. *TechTrends*, *67*(6), 878–884. https://doi.org/10.1007/s11528-023-00914-1

Fauzi, F., Tuhuteru, L., Sampe, F., Ausat, A., & Hatta, H. R. (2023). Analysing the Role of ChatGPT in Improving Student Productivity in Higher Education. *Journal on Education*, *5*, 14886–14891. https://doi.org/10.31004/joe.v5i4.2563

Fissore, C., Floris, F., Fradiante, V., Marchisio Conte, M., & Sacchet, M. (2024). From Theory to Training: Exploring Teachers' Attitudes Towards Artificial Intelligence in Education: *Proceedings of the 16th International Conference on Computer Supported Education*, 118–127. https://doi.org/10.5220/0012734700003693

Galindo-Domínguez, H., Delgado, N., Campo, L., & Losada, D. (2024). Relationship between teachers' digital competence and attitudes towards artificial intelligence in education. *International Journal of Educational Research*, *126*, 102381. https://doi.org/10.1016/j.ijer.2024.102381

Galindo-Domínguez, H., Delgado, N., Losada, D., & Etxabe, J.-M. (2024). An analysis of the use of artificial intelligence in education in Spain: The in-service teacher's perspective. *Journal of Digital Learning in Teacher Education*, *40*(1), 41–56. https://doi.org/10.1080/21532974.2023.2284726

Higgs, J. M., & Stornaiuolo, A. (2024). Being Human in the Age of Generative AI: Young People's Ethical Concerns about Writing and Living with Machines. *Reading Research Quarterly*, *59*(4), 632–650. https://doi.org/10.1002/rrq.552

Hornberger, M., Bewersdorff, A., & Nerdel, C. (2023). What do university students know about Artificial Intelligence? Development and validation of an AI literacy test. *Computers and Education: Artificial Intelligence*, *5*, 100165. https://doi.org/10.1016/j.caeai.2023.100165

Kaplan-Rakowski, R., Grotewold, K., Hartwick, P., & Papin, K. (2023). Generative AI and Teachers' Perspectives on Its Implementation in Education. *Journal of Interactive Learning Research*, *34*, 313–338.

Karan, B., & Angadi, G. R. (2023). Potential Risks of Artificial Intelligence Integration into School Education: A Systematic Review. *Bulletin of Science, Technology & Society*, *43*(3–4), 67–85. https://doi.org/10.1177/02704676231224705

Kasneci, E., Sessler, K., Küchemann, S., Bannert, M., Dementieva, D., Fischer, F., Gasser, U., Groh, G., Günnemann, S., Hüllermeier, E., Krusche, S., Kutyniok, G., Michaeli, T., Nerdel, C., Pfeffer, J., Poquet, O., Sailer, M., Schmidt, A., Seidel, T., … Kasneci, G. (2023). ChatGPT for good? On opportunities and challenges of large language models for education. *Learning and Individual Differences*, *103*, 102274. https://doi.org/10.1016/j.lindif.2023.102274

Kotchetkov, M., & Trevor, C. (2025). ChatGPT: Hero or Villain? Comparative Evaluation by Canadian High School Students and Teachers. *Canadian Journal of Educational and Social Studies*, *5*(3), 91–108. https://doi.org/10.53103/cjess.v5i3.364

Krupp, L., Steinert, S., Kiefer-Emmanouilidis, M., Avila, K., Lukowicz, P., Kuhn, J., Küchemann, S., & Karolus, J. (2024). *Unreflected Acceptance – Investigating the Negative Consequences of ChatGPT-Assisted Problem Solving in Physics Education*. https://doi.org/10.3233/FAIA240195

Küchemann, S., Avila, K., Dinc, Y., Hortmann, C., Revenga Lozano, N., Ruf, V., Stausberg, N., Steinert, S., Fischer, F., Fischer, M., Kasneci, E., Kasneci, G., Kuhr, T., Kutyniok, G., Malone, S., Sailer, M., Schmidt, A., Stadler, M., Weller, J., & Kuhn, J. (2024). *Are Large Multimodal Foundation Models all we need? On Opportunities and Challenges of these Models in Education*. https://doi.org/10.35542/osf.io/n7dvf

Küchemann, S., Steinert, S., Kuhn, J., Avila, K., & Ruzika, S. (2024a). Large language models—Valuable tools that require a sensitive integration into teaching and learning physics. *The Physics Teacher*, *62*, 400–402. https://doi.org/10.1119/5.0212374

Küchemann, S., Steinert, S., Kuhn, J., Avila, K., & Ruzika, S. (2024b). Large language models—Valuable tools that require a sensitive integration into teaching and learning physics. *The Physics Teacher*, *62*, 400–402. https://doi.org/10.1119/5.0212374

Kuleto, V., Ilić, M. P., Bucea-Manea-Țoniş, R., Ciocodeică, D.-F., Mihălcescu, H., & Mindrescu, V. (2022). The Attitudes of K–12 Schools' Teachers in Serbia towards the Potential of Artificial Intelligence. *Sustainability*, *14*(14), Article 14. https://doi.org/10.3390/su14148636

Lademann, J., Henze, J., & Becker-Genschow, S. (2025). Augmenting learning environments using AI custom chatbots: Effects on learning performance, cognitive load, and affective variables. *Physical Review Physics Education Research*, *21*(1), 010147. https://doi.org/10.1103/PhysRevPhysEducRes.21.010147

Laupichler, M., Aster, A., Schirch, J., & Raupach, T. (2022). Artificial intelligence literacy in higher and adult education: A scoping literature review. *Computers and Education: Artificial Intelligence*, *3*, 100101. https://doi.org/10.1016/j.caeai.2022.100101

Lo, C. K. (2023). What Is the Impact of ChatGPT on Education? A Rapid Review of the Literature. *Education Sciences*, *13*(4), Article 4. https://doi.org/10.3390/educsci13040410

Long, D., & Magerko, B. (2020). What is AI Literacy? Competencies and Design Considerations. *Proceedings of the 2020 CHI Conference on Human Factors in Computing Systems*, 1–16. https://doi.org/10.1145/3313831.3376727

Moura, A., & Carvalho, A. A. A. (2024). Teachers' perceptions of the use of artificial intelligence in the classroom. In O. Titrek, C. S. De Reis, & J. G. Puerta (Eds.), *International Conference on Lifelong Education and Leadership for All (ICLEL 2023)* (Vol. 17, pp. 140–150). Atlantis Press International BV. https://doi.org/10.2991/978-94-6463-380-1_13

Nazaretsky, T., Ariely, M., Cukurova, M., & Alexandron, G. (2022). Teachers' trust in AI-powered educational technology and a professional development program to improve it. *British Journal of Educational Technology*, *53*(4), 914–931. https://doi.org/10.1111/bjet.13232

Neumann, K., Kuhn, J., & Drachsler, H. (2024). Generative Künstliche Intelligenz in Unterricht und Unterrichtsforschung – Chancen und Herausforderungen [Generative artificial intelligence in teaching and education research - opportunities and challenges]. *Unterrichtswissenschaft*, *52*(2), 227–237. https://doi.org/10.1007/s42010-024-00212-6

Ng, D. T. K., Chan, E. K. C., & Lo, C. K. (2025). Opportunities, challenges and school strategies for integrating generative AI in education. *Computers and Education: Artificial Intelligence*, *8*, 100373. https://doi.org/10.1016/j.caeai.2025.100373

OECD. (2015). *Students, Computers and Learning: Making the Connection*. OECD. https://doi.org/10.1787/9789264239555-en

Popenici, S. A. D., & Kerr, S. (2017). Exploring the impact of artificial intelligence on teaching and learning in higher education. *Research and Practice in Technology Enhanced Learning*, *12*(1), 22. https://doi.org/10.1186/s41039-017-0062-8

R Core Team. (2024). *R: A Language and Environment for Statistical Computing* (Version 4.4.0) [Computer software]. https://www.R-project.org

Revelle, W. (2025). *psych: Procedures for Psychological, Psychometric, and Personality Research* (p. 2.5.3) [Dataset]. https://doi.org/10.32614/CRAN.package.psych

Rienties, B. (2014). Understanding academics' resistance towards (online) student evaluation. *Assessment & Evaluation in Higher Education*, *39*(8), 987–1001. https://doi.org/10.1080/02602938.2014.880777

Sanusi, I. T., Oyelere, S. S., Vartiainen, H., Suhonen, J., & Tukiainen, M. (2023). A systematic review of teaching and learning machine learning in K-12 education. *Education and Information Technologies*, *28*(5), 5967–5997. https://doi.org/10.1007/s10639-022-11416-7

Seßler, K., Xiang, T., Bogenrieder, L., & Kasneci, E. (2023). *PEER: Empowering Writing with Large Language Models* (pp. 755–761). https://doi.org/10.1007/978-3-031-42682-7_73

Stacey, M., McGrath-Champ, S., & Wilson, R. (2023). Teacher attributions of workload increase in public sector schools: Reflections on change and policy development. *Journal of Educational Change*, *24*(4), 971–993. https://doi.org/10.1007/s10833-022-09476-0

Sugar, W., Crawley, F., & Fine, B. (2004). Examining Teachers' Decisions To Adopt New Technology. *Educational Technology & Society*, *7*, 201–213.

UNESCO. (2024). *AI competency framework for teachers*. https://doi.org/10.54675/ZJTE2084

Vodafone Stiftung. (2025). *KI an europäischen Schulen [AI in European schools]*. https://www.vodafone-stiftung.de/europaeische-schuelerstudie-kuenstliche-intelligenz/

Wang, T., & Cheng, E. C. K. (2021). An investigation of barriers to Hong Kong K-12 schools incorporating Artificial Intelligence in education. *Computers and Education: Artificial Intelligence*, *2*, 100031. https://doi.org/10.1016/j.caeai.2021.100031

Wang, X., Li, L., Tan, S. C., Yang, L., & Lei, J. (2023). Preparing for AI-enhanced education: Conceptualizing and empirically examining teachers' AI readiness. *Computers in Human Behavior*, *146*, 107798. https://doi.org/10.1016/j.chb.2023.107798

Wollny, S., Schneider, J., Di Mitri, D., Weidlich, J., Rittberger, M., & Drachsler, H. (2021). Are We There Yet? - A Systematic Literature Review on Chatbots in Education. *Frontiers in Artificial Intelligence*, *4*. https://doi.org/10.3389/frai.2021.654924

Wu, R., & Yu, Z. (2024). Do AI chatbots improve students learning outcomes? Evidence from a meta-analysis. *British Journal of Educational Technology*, *55*(1), 10–33. https://doi.org/10.1111/bjet.13334

Yang, T.-C., & Chen, J.-H. (2023). Pre-service teachers' perceptions and intentions regarding the use of chatbots through statistical and lag sequential analysis. *Computers and Education: Artificial Intelligence*, *4*, 100119. https://doi.org/10.1016/j.caeai.2022.100119

Zhai, X. (2023a). *ChatGPT and AI: The Game Changer for Education* (SSRN Scholarly Paper No. 4389098). Social Science Research Network. https://papers.ssrn.com/abstract=4389098

Zhai, X. (2023b). ChatGPT for Next Generation Science Learning. *XRDS*, *29*(3), 42–46. https://doi.org/10.1145/3589649

Zheng, Z., Sun, Y., Song, X., Zhu, H., & Xiong, H. (2023). Generative Learning Plan Recommendation for Employees: A Performance-aware Reinforcement Learning Approach. *Proceedings of the 17th ACM Conference on Recommender Systems*, 443–454. https://doi.org/10.1145/3604915.3608795